\documentclass{article}

\usepackage{amsmath,amstext,amsthm,enumerate,graphicx}

\def\a{{\alpha}}

\def\b{{\beta}}
\def\w{{\omega}}

\def\h{\hbar}

\def\"{{\guillemotleft}}

\def\"{{\guillemotright}}

\numberwithin{equation}{section}

\theoremstyle{plain}

\newtheorem{req}{Remark}

\newtheorem{prop}{Proposition}[section]

\theoremstyle{definition}

\begin{document}

\title{Superintegrability with third-order integrals in quantum and classical mechanics}

 \author{Simon Gravel\\\textit{\small D\'epartement de physique et Centre de recherche math\'ematiques}\\
 \textit{\small Universit\'e de Montr\'eal, C.P.\-6128, Succursale Centre-Ville} \\
 \textit{\small Montr\'eal, Qu\'ebec} \\ \textit{\small H3C 3J7} \\ \textit{\small Canada}\\\small graves@magellan.umontreal.ca\\
 \and Pavel Winternitz\\\textit{\small D\'epartement de math\'ematiques et de statistique et Centre de recherche math\'ematiques}\\
 \textit{\small Universit\'e de Montr\'eal, C.P.\-6128, Succursale Centre-Ville} \\
 \textit{\small Montr\'eal, Qu\'ebec} \\ \textit{\small H3C 3J7} \\ \textit{\small Canada}\\\small wintern@crm.umontreal.ca}

 \date{June 26, 2002}

 %\subjclass{}

\maketitle

\begin{abstract}

We consider here the coexistence of first- and third-order
integrals of motion in two dimensional classical and quantum
mechanics. We find explicitly all potentials that admit such
integrals, and all their integrals. Quantum superintegrable
systems are found that have no classical analog, i.e. the
potentials are proportional to $\h^2$, so their classical limit is
free motion.

\end{abstract}

\pagebreak

 \maketitle

\section{Introduction}

\label{sintro}

In classical mechanics, an n-dimensional Hamiltonian system is
called Liouville integrable if it allows $n$ functionally
independent integrals of motion in involution (including the
Hamiltonian), that is

\begin{equation}
\begin{split}
\{H,X_i\}&=0,\\
\{X_i,X_j\}&=0,  \forall i,j.
\end{split}
\end{equation}

The Hamiltonian $H=H(x_1,...x_n,p_1,...,p_n)$ and the integrals of
motion $X_i=X_i(x_1,...x_n,p_1,...,p_n)$ must be well defined
functions on phase space~(\cite{Ar,Go}). The system is
superintegrable if it allows more than $n$ functionally
independent integrals, $n$ of them in involution. The best known
superintegrable systems in $n$ dimensions are the harmonic
oscillator $V=\omega r^2$ and the Coulomb potential
$V=\frac{\a}{r}$, both of them allowing $2 n-1$ independent
integrals of motion, the maximal number possible for an
interacting system. Bertrand's theorem (\cite{Ar, Be}) tells us
that these are the only rotationally invariant systems for which
all finite trajectories are closed, a fact intimately related to
their maximal superintegrability.

In quantum mechanics, a Hamiltonian system is said to be
integrable if there exists a set $\{X_i\}$ of $n$ well defined,
algebraically independent operators (including the Hamiltonian)
that commute pairwise. It is superintegrable if it possesses
further independent operators, $\{Y_j\}$ that commute with the
Hamiltonian. The $Y_j$ do not  necessarily commute with each
other, nor with the $X_i$.

The definition of the independence of quantum operators is not
unique, and this may give rise to different types of quantum
superintegrability.
%We may say operators are algebraically
%independent if there isn't a polynomial in the Jordan algebra of
%the operators that vanishes. In two dimensions this is equivalent to asking that no polynomial $P(x_1,...,x_n)$ exist such that $P(x_1,...,x_n)+P(x_1,...,x_n)\neq 0$ and$P(X_1,...,X_n)+P(X_1,...,X_n)=0$, but these definition are no more equivalent in more than two dimensions, according to Cohn's theorem[[citer Jacobson?]].

A good working definition, which may be appropriate for
applications in quantum mechanics, soliton theory and for instance
in the study of the Huygens principle, is that operators are
considered independent unless one of them can be expressed as a
polynomial in the others (\cite{BC,CV,Ht1,Ht3}). The fact that
commuting operators can be useful even if they are functionally
dependent in the classical limit was clearly demonstrated by
Hietarinta (\cite{Ht1,Ht2,HG,Ht3}). This definition is in itself
not quite satisfactory since it ignores more general polynomial or
functional relations between integrals. This may lead to important
differences between classical and quantum integrability. Moreover,
it is not appropriate for nonpolynomial integrals. Finding an
appropriate and rigorous definition of the independence of quantum
operators is not an easy problem, but it is worth investigating as
wrong or ambiguous definitions may give rise to incorrect results.
For a discussion of related problems, see e.g. \cite{HG} and
\cite{We}

Previous systematic searches for superintegrable systems
concentrated on integrals of motion of at most second order in
momenta (\cite{Ev1,Ev2,Ev3,FMSUW,MSVW,WSF}). This ``quadratic
superintegrability'' has been shown to be related to
multiseparability of the Schroedinger or Hamilton-Jacobi
equations. More recently, it was related to generalized symmetries
(\cite{STW}) and exact solvability (\cite{TTW}).

Quadratic superintegrability has been considered in spaces of
nonzero constant curvature (\cite{KMP,RS}) and of nonconstant
curvature (\cite{KKW}) . For superintegrable systems in $n$
dimensions see (\cite{RW}).

 The purpose of this article is to start a systematic
search for superintegrable systems with higher order integrals of
motion. We consider a two-dimensional real Euclidian space with a
one-particle Hamiltonian;

\begin{equation*}
H=\frac{1}{2}\left(p_x^2+p_y^2\right)+V(x,y).
\end{equation*}

We request the existence of two additional integrals of motion,
one of first order in the momenta and the other of third order.

The classical and quantum mechanical cases will be treated
separately. When second order integrals of motion are considered,
classical and quantum integrable and superintegrable potentials
coincide. For third order integrals this is no longer the case (as
was pointed out by Hietarinta in \cite{Ht1}). For integrable
systems with third or higher order integrals in classical
mechanics, see also \cite{Dr,FL1,FL2,GKM,Ho,KR,Ra}

\section{Conditions for the existence of a third order invariant in classical mechanics}

\label{classique}

We are looking for a classical integral of motion that is a
polynomial in the momenta with coefficients depending on the
spatial coordinates, i.e.

\begin{equation*}
X=\sum_{j,k} f_{jk}(x,y)p_1^jp_2^k,
\end{equation*}

\noindent that Poisson-commutes with the Hamiltonian;
\begin{equation}\label{poisson}
\begin{split}
0&=\{H,X\},\\
H&=\frac{p_1^2+p_2^2}{2}+V(x,y).
\end{split}
\end{equation}

We can simplify our search by using the fact that equation
\eqref{poisson} implies that $X$ is a constant over any
trajectory:
\begin{equation}\label{const}
\frac{dX}{dt}=\frac{\partial X}{\partial q_i}\dot
q_i+\frac{\partial X}{\partial p_i}\dot p_i=0
\end{equation}

\noindent with
\begin{equation}
\begin{split}\label{hamilton}
\dot p_i &=-V_{q_i}(q_1,q_2),\\
\dot q_i &=p_i.
\end{split}
\end{equation}

If we write explicitly $X$ in \eqref{const}, we find

\begin{equation}\label{enveloppe}
  \sum_{j+k=1}^n \left(\frac{\partial f_{jk}}{\partial x}p_1^{j+1}p_2^k+\frac{\partial f_{jk}}{\partial
  y}p_1^{j}p_2^{k+1}-f_{jk}V_x j p_1^{j-1}p_2^k-f_{jk}V_y k p_1^j
  p_2^{k-1}\right)=0.
\end{equation}

Since the monomials $p_1^a p_2^b$'s form a basis, the coefficients
for each $(a,b)$ must vanish separately, thus \eqref{enveloppe}
gives relations between the $f_{ij}$ with odd and even $i+j$
separately. If we are looking for an integral of odd (even) degree
in the momenta, the even (odd) terms will play no role and we can
without loss of generality consider only integrals that have terms
only of odd (even) parity. Moreover, we may notice, in
\eqref{enveloppe}, that the terms of leading order in the $p_i$'s
imply a relation independent of $V$ between the $f_{i,j}$ with
$i+j=n$. This allows us to find immediately the form of the
leading order terms, so the integral of motion in the third-order
case takes the form

\begin{equation}
\begin{split}
X & = \sum_{i+j+k=3} A_{ijk} p_1^i p_2^j L^k +g_1(x,y) p_1 + g_2(x,y) p_2,\\
L & = xp_2-yp_1.
\end{split}
\label{constante}
\end{equation}

\noindent where the $A_{ijk}$ are arbitrary real constants.

The requirement $\frac{dX}{dt}=0$ and the Hamilton equations
(\eqref{hamilton}) yield four equations

\begin{eqnarray}
0 =&& g_1V_x +g_2V_y,\label{eqclass1}\\
(g_1)_x =&& 3 f_1(y) V_x +f_2(x,y) V_y,\label{eqclass2}\\
(g_2)_y =&& f_3(x,y) V_x +3 f_4(x) V_y,\label{eqclass3}\\
(g_1)_y +(g_2)_x =&& 2 \left(f_2(x,y) V_x+f_3(x,y) V_y \right),
\label{eqclass4}
\end{eqnarray}

\noindent where

\begin{align*}
f_1(y) &=- A_{300} y^3 +A_{210}y^2-A_{120}y +A_{030},\\
f_2(x,y) &=3A_{300}xy^2- 2A_{210}xy +A_{201}y^2 +A_{120}x-A_{111}y+A_{021},\\
f_3(x,y) &=-3A_{300}x^2y+A_{210}x^2 -2 A_{201}xy +A_{111} x-A_{102}y+A_{012},\\
f_4(x) &=A_{300} x^3+A_{201} x^2+A_{102} x +A_{003},.
\end{align*}

Requiring that equations \eqref{eqclass2},\eqref{eqclass3}, and
\eqref{eqclass4} be compatible, we obtain a linear compatibility
condition for the potential, namely

\begin{equation}
\label{compatlin}
\begin{split}
0=&-f_3 V_{xxx}+\left(2 f_2-3 f_4\right) V_{xxy}+\left(-3 f_1+2 f_3\right) V_{xyy}-f_2 V_{yyy}\\
&+2\left( f_{2y}- f_{3x}\right) V_{xx}+2\left(-3f_{1y}+f_{2x}+f_{3y}-3f_{4x}\right)V_{xy}+2\left(-f_{2y}+f_{3x}\right)V_{yy}\\
&+\left(-3f_{1yy}+2f_{2xy}-f_{3xx}\right)
V_x+\left(-f_{2yy}+2f_{3xy}-3f_{4xx}\right) V_y.
\end{split}
\end{equation}

Requiring that all four equations \eqref{eqclass1}
\eqref{eqclass2},\eqref{eqclass3},\eqref{eqclass4} be compatible,
we obtain further third-order equations for the potential, this
time nonlinear ones. They are limit case (for $\h \rightarrow 0$)
of the corresponding quantum compatibility conditions
\eqref{compatnl1} to \eqref{compatnl3} given below.

These conditions,  together with \eqref{compatlin}, form an
overdetermined system for the potential $V(x,y)$. The solution
space will hence be rather restricted. Indeed, in 1935, Drach
(\cite{Dr}) posed the problem of finding classical Hamiltonian
systems with one third-order integral. In a complex Euclidian
space $E_2(\bf{C})$ he found $10$ such potentials, each one
depending on arbitrary constants, not however on arbitrary
functions. We recall that in the case of second order integrals,
one obtains four families of potentials, each of them depending on
two arbitrary functions of one variable (\cite{FMSUW,WSF}). They
are the four most general potentials that allow separation of
variables in cartesian, polar, parabolic and elliptic coordinates,
respectively.

\section{Conditions for the existence of a third order invariant in quantum mechanics}

\label{condition}

Here we are interested in the existence of third-order operators,i.e.

\begin{equation*}
\begin{split}
X = \sum_{i+j=0}^{3} P_{ij}(x,y)p_1^ip_2^j,\\
p_1=-i \hbar \partial_x,~~p_2=-i \hbar \partial_y,
\end{split}
\end{equation*}

\noindent  that commute with the Hamiltonian. An equivalent way of
writing this operator is

\begin{equation*}
X = \sum_{i+j=0}^{3} \{P_{ij}(x,y),p_1^ip_2^j\}.
\end{equation*}

Here the bracket means the anticommutator:

$$\{f,p_1^jp_2^k\}=f p_1^jp_2^k+p_1^jp_2^kf.$$

Each of these anticommutators can be expressed as

$$\{f,p_1^jp_2^k\}^+ + i \{f,p_1^jp_2^k\}^- = \{\Re{e[f]},p_1^jp_2^k\} + i \{\Im{m[f]},p_1^jp_2^k\}.$$

\noindent Hence we can write the operator $X$ in the form

$$ X=X^+ +i X^-,$$

\noindent where $X^+$ and $X^-$ are self-adjoint operators. As the
Hamiltonian itself is self-adjoint, $X^\dag=X^+-iX^-$ must also
commute, as well as $X^+$ and $X^-$. These two last operators
commute under the same conditions, so we may restrict our search
without loss of generality to self-adjoint operators. This turns
out to be quite useful in view of the following result

\begin{prop}
\label{parite}

For each self-adjoint integral of motion of order $n$, there
exists one integral of order $n$ with definite parity,i.e.

\begin{equation}
X_n=\sum_{j=0}^{[\frac{n}{2}]}\sum_{k=0}^{n-2j}
\{P_{n-2j,k}(x,y),p_1^kp_2^{n-2j-k}\},
 \label{opsimple}
\end{equation}

\noindent where $P$ is a real function.
\end{prop}
\begin{proof}
This is simply due to the fact that we have a real Hamiltonian and
a purely imaginary momentum operator, so terms of even order,
which are real, must commute independently of the terms of odd
order, which are purely imaginary.

\end{proof}

In the case $n=3$ we restrict ourselves to third-order integrals
of the form
\begin{equation*}
X_3=\sum_{i+j=3}
\{f_{ij}(x,y),p_1^jp_2^k\}+\{g_1(x,y),p_1\}+\{g_2(x,y),p_2\}.
\end{equation*}

Requesting

\begin{equation*}
\begin{split}
0&=\left[ H,X \right],\\
H&=\frac{1}{2m} \left(p_1^2+p_2^2\right)+V(x,y),
\end{split}
\end{equation*}

\noindent we find a set of 15 differential equations, of which the
first nine can be explicitly solved to give

\begin{equation}
X=\sum_{\substack{i,j,k\\i+j+k=3}}A_{ijk}
\{L_3^i,p_1^jp_2^k\}+\{g_1(x,y), p_1\}+\{g_2(x,y), p_2\}.
\label{opnorm}
\end{equation}

\noindent where the $A_{ijk}$ are arbitrary real constants. So far
this is similar to the classical case.

\begin{req}
The argument used in demonstrating proposition \ref{parite} can be
generalized to any expression involving the anticommutators of
self-adjoint operators homogeneous in the $p_i$'s, for example to
terms of the form $\{L_3^i,p_1^jp_2^k\}$, as long as the
coefficients of the $p_i$'s are real.

\end{req}

We could get rid of the $\hbar$ and $m$ factors by a dilation of
the undetermined functions,
\begin{equation*}
\begin{split}
V(x,y)&=\frac{\hbar^2}{2 m}\tilde{V}(x,y),\\
g_1(x,y)&=\hbar^2 g_1'(x,y),\\
g_2(x,y)&=\hbar^2 g_2'(x,y).
\end{split}
\end{equation*}

This is equivalent to setting $\hbar$ and $m$ equal to one, which
we could do, but we prefer to keep track of the dependence on
$\hbar$ (while setting $m=1$), in order to see the classical
limit.

 We are left with a set of 6 equations, two
of which are consequences of the other four, so that, as in the
classical case, we have to solve four equations:

\begin{align}
\begin{split}
0 &= g_1V_x +g_2V_y-\frac{\h^2}{4}\Big(f_1 V_{xxx}
+f_2V_{xxy}+f_3V_{xyy}+f_4V_{yyy}\\&~~~~~~+8A_{300}(xV_y-yV_x)
+2\left(A_{210}V_x +A_{201}V_y\right)\Big),
\end{split}\label{eqquant1}\\
(g_1)_x &= 3 f_1(y) V_x +f_2(x,y) V_y\equiv h_1,\label{eqquant2}\\
(g_2)_y &= f_3(x,y) V_x +3 f_4(x) V_y\equiv h_2,\label{eqquant3}\\
(g_1)_y +(g_2)_x &= 2 \left(f_2(x,y) V_x+f_3(x,y) V_y
\right)\equiv h_3. \label{eqquant4}
\end{align}

Equations \eqref{eqquant2} to \eqref{eqquant4} are the same as in
the classical case, however equation \eqref{eqquant1} differs from
equation \eqref{eqclass1} by the terms proportional to $\h^2$.
Both in the classical and quantum cases we can eliminate $g_1$ and
$g_2$ and obtain compatibility conditions for the potentials.

We shall write these in a unified manner for both cases. One such
compatibility condition is the third-order linear equation
\eqref{compatlin}. To write three more conditions we introduce the
notation

\begin{equation*}
\begin{split}
\phi_1 &=\frac{V_y}{V_x},\\
\phi_2&=-\h^2\frac{\Big(f_1 V_{xxx}
+f_2V_{xxy}+f_3V_{xyy}+f_4V_{yyy}+8A_{300}(xV_y-yV_x)
+2\left(A_{210}V_x +A_{201}V_y\right)\Big)}{4 V_x}.
\end{split}
\end{equation*}

\noindent and use $h_1$,$h_2$ and $h_3$ introduced above. In the
classical case we have $\phi_2=0$. The three (independent)
nonlinear compatibility conditions are

\begin{equation}\label{compatnl1}
-\phi_{2x} +\left(\frac{\phi_1\left(h_3 \phi_1 +h_2
\phi_1^2+\phi_1\phi_{2y}+\phi_{2x}
+h_4\right)}{\phi_{1x}+\phi_1\phi_{1y}}\right)_x= h_4,
\end{equation}

\begin{equation}\label{compatnl2}
\left(\frac{\phi_1^2 h_5+ \phi_1\phi_{2y} +\phi_1 h_6 +
\phi_{2x}+h_4}{\phi_{1x}+\phi_1 \phi_{1y}}\right)_y =-h_5,
\end{equation}

\begin{equation}\label{compatnl3}
\begin{split}
h_4&\left(\phi_{1xy}+\phi_{1y}^2\right)+h_5\left(
\phi_1^2\phi_{1xy}-\phi_{1x}^2-2\phi_1\phi_{1x}\phi_{1y}\right)
\\+h_6&\left(\phi_1\phi_{1xy}-\phi_{1x}\phi_{1y}\right)
-\left(h_{4y}+\phi_1h_{5x}\right)\left(\phi_{1x}+\phi_1\phi_{1y}\right)\\&=-\phi_{2x}\left(\phi_{1xy}+\phi_{1y}^2\right)+\phi_{2y}
\left(\phi_{1x}\phi_{1y}-\phi_{1xy}\phi_1\right)+\phi_{2xy}
\left(\phi_{1x}+\phi_1 \phi_{1y}\right).
\end{split}
\end{equation}

In the quantum case these are fifth order equations for the
potential. They can be used to express $\phi_{2xy}$, $\phi_{2xx}$
and $\phi_{2yy}$in terms of $\phi_{2x}$, $\phi_{2y}$ and $\phi_2$.
In the classical case we set $\phi_2=0$, but the equations remain
independent. They can be used to determine $\phi_{1xy}$,
$\phi_{1xx}$ and $\phi_{1yy}$in terms of $\phi_{1x}$, $\phi_{1y}$
and $\phi_1$. The nonlinear equations for V(x,y) are third order
ones in the classical case.

In deriving these equations we have assumed
\begin{equation}
\begin{split}
\phi_{1x} +\phi_1 \phi_{1y}\neq 0,\\
\phi_{1xy}\phi_{1}-\phi_{1x}\phi_{1y} \neq 0.
\end{split}
\end{equation}

The cases when the above conditions do not hold must be considered
separately. This will actually be the case for potentials
considered in this article.

 We also mention the interesting
fact, already noticed by Hietarinta (\cite{Ht2}),that a classical
integrable potential is also quantum integrable, if and only if it
respects the compatibility condition

\begin{equation}
f_1 V_{xxx} +f_2V_{xxy}+f_3V_{xyy}+f_4V_{yyy}+8A_{300}(x V_y
-yV_x) +2\left(A_{210}V_x +A_{201}V_y\right)=0.
 \label{condnouv}
\end{equation}

 In that case the equations are invariant
under a simultaneous dilation of the potential and the $g_i$'s.
Thus any potential that is a solution to both \eqref{eqclass1} to
\eqref{eqclass4} and \eqref{condnouv} can be multiplied by an
arbitrary factor, which can be used to "absorb" the $\hbar^2$
factor so the solution does not vanish in the classical limit.

Even if a classical superintegrable potential does not satisfy
this relation, there could exist corresponding quantum
superintegrable systems. In that case, though, the equations are
not invariant under a dilation of the potential as in the previous
case, so terms that do not satisfy both \eqref{condnouv} and
 \eqref{eqclass1} to \eqref{eqclass4}  must be proportional to $\hbar^2$ and vanish
in the classical limit.

We will show that condition \eqref{condnouv} cannot be the
consequence of equations \eqref{eqquant2}, \eqref{eqquant3} and
\eqref{eqquant4} for in that case the classical and quantum
integrable potentials would be the same.

\section{Superintegrable systems with one third order and one first order integral}

\subsection{Integral of first-order}

A potential $V(x,y)$ allows an integral that is of first order in
the momenta if and only if it is invariant under either rotations
or translations. Thus the potential must satisfy
$$aL_3 V+bp_1 V + cp_2 V=0.$$

Without loss of generality, we can take the potential and first
order integral to be one of the following:

\begin{itemize}
\item $a\neq 0$:\space \space \space \space \space \space \space \space \space \space \space \space \space $V=V(r)$,\space \space \space \space $X=L_3$
\item $a=0$, $b^2+c^2\neq 0$: $V=V(x)$,\space \space \space \space$X=p_2$.
\end{itemize}

\subsection{Quantum and classical superintegrable potentials invariant under rotations.}

Compatibility conditions obtained from equations \eqref{eqclass1}
to \eqref{eqclass4} or \eqref{eqquant1} to \eqref{eqquant4} leave
us with only two possibilities, namely

\begin{equation*}
\begin{split}
V=\frac{\alpha}{r},\\
V=\omega^2 r^2
\end{split}
\end{equation*}

The Coulomb potential and the harmonic oscillator, which are the
best-known superintegrable potentials in any dimension. In
addition to angular momentum $L_3$, the Coulomb potential in $E_2$
allows two second order integrals, namely the components of the
Laplace-Runge-Lenz vector:

\begin{equation*}
\begin{split}
X_1^C=\{L_3,p_1\}-\frac{2 \alpha y}{r};\\
X_2^C=\{L_3,p_2\}+\frac{2 \alpha x}{r}.
\end{split}
\end{equation*}

The harmonic oscillator, in addition to angular momentum, allows
two second order integrals which are the components of a
quadrupole tensor;

\begin{equation*}
\begin{split}
X_1^h=-\frac{1}{2} p_1^2+\frac{1}{2} p_2^2+\omega^2 x^2-\omega^2 y^2;\\
X_2^h=-p_1p_2 + 2 \omega^2xy.
\end{split}
\end{equation*}

Commuting (or Poisson commuting) second order integrals, we in
general find third order integrals.

The third order integrals obtained for these potentials are indeed
direct consequences of integrals at order one and two.

%[[We notice the fact that even if the first and second-order
%symmetries are equivalent in classical and quantum mechanics, the
%quantum third-order operators can be modified by a factor of
%$\hbar^2$. For example

%\begin{equation}
%\{X_1^C,L_3\}= \{L_3^2,p_1\}+\{\frac{\alpha y^2}{r}-\frac{\hbar^2}{4},p_1\}-\{\frac{\alpha xy}{r},p_2\}
%\end{equation}

%is the quantum equivalent to the classical integral
%\begin{equation}
%L_3^2p_1+\frac{\alpha y^2}{r}p_1-\frac{\alpha xy}{r}p_2.
%\end{equation}]]

%calcule dans vrclass

\subsection{Classical superintegrable potentials invariant under translation.}

In the classical case, the remaining equations are readily solved.
If we set $V_y=0$, equations \eqref{eqclass1} to \eqref{eqclass4}
simplify to

\begin{gather*}
0= g_1\\
0=A_{300}=A_{210}=A_{120}=A_{030}\\
(g_2)_y=f_3(x,y) V_x\\
(g_2)_x=2 f_2(x,y) V_x
\end{gather*}

We can at once set $A_{021}$ and $A_{003}$ to 0, for they
correspond to trivial constants of motion, $p_2^3$ and $H p_2$,
that can be subtracted from the constant \eqref{constante}.

The compatibility condition between the two last equations forces
one of the three following conditions to be satisfied (up to a
translation in x).

\begin{gather}
V= a x \label{ax}\\
V= \frac{a}{x^2} \label{asx2}\\
 A_{201}=A_{111}=A_{102}=A_{012}=0
\end{gather}

The first two potentials correspond to superintegrable systems
that have one first and at least one second order integral. Their
third order integrals can be obtained by commutation of these
lower-order ones.

The last conditions forbids the existence of a nontrivial
third-order commuting operator for any other potentials than
\eqref{ax} and \eqref{asx2}.

\subsection{Quantum superintegrable potentials invariant under translation.}

Here the situation is more interesting. Equations \eqref{eqquant1}
to \eqref{eqquant4} reduce to
\begin{align}
0 & = g_1V_x-\frac{\h^2}{4}\Big(f_1 V_{xxx}
-8y A_{300}V_x +2 A_{210}V_x\Big)\label{eqx1}\\
(g_1)_x &= 3 f_1(y) V_x\label{eqx2}\\
(g_2)_y &= f_3(x,y) V_x\label{eqx3}\\
(g_1)_y +(g_2)_x &= 2 \left(f_2(x,y) V_x \right) \label{eqx4}
\end{align}

The linear compatibility condition leads to two equations (since
coefficients of $y^0$ and of $y^1$ must vanish separately), namely
\begin{equation}
\label{compatx}
\begin{split}
0&=(A_{210}x^2+A_{111}x+A_{012})V_{xxx} +4 (2A_{210} x+A_{111})V_{xx}+12A_{210}V_x;\\
0&=(3 A_{300}x^2+2A_{201}x+A_{102})V_{xxx} +4 (6A_{300} x+2
A_{201})V_{xx}+36A_{300}V_x.
\end{split}
\end{equation}

The two equations are similar and easy to solve, but it turns out
their only solutions that also satisfy \eqref{eqx1} to
\eqref{eqx4} are again the potentials $V=ax$ and $V=a/x^2$. Their
third-order integrals in general are direct consequences of
lower-order commuting operators, that is they can be obtained by
commuting their second order integrals. In the $V=\frac{a}{x^2}$
case, we find three third order integrals,
\begin{equation*}
\begin{split}
X_1 &= \{L_3^2,p_2\}+ a\{2 \frac{y^2}{x^2},p_2\}\\
X_2 &= \{L_3,p_1 p_2\}- a\{4 \frac{y}{x^2},p_2\}\\
X_3 &= p_1^2p_2- a\{ \frac{4}{x^2},p_2\}
\end{split}
\end{equation*}
The integrals $X_2$ and $X_3$ can be obtained by commuting $X_1$
with the first-order integral $p_2$.

In the particular case $V=\frac{\hbar^2}{x^2}$, we find four more
integrals, again related to each other by commutation with $p_2$;

\begin{equation*}
\begin{split}
X_4 &=
L_3^3+\frac{\hbar^2}{2}\{\frac{6y^2}{x}+2x,p_2\}+\frac{\hbar^2}{2}\{\frac{-3y^3}{x^2}-2y,p_1\}\\
X_5 &=\{L_3^2,p_1\}-\hbar^2\{\frac{4y}{x},p_2\}+\frac{\hbar^2}{2}\{ \frac{6 y^2}{x^2}+1,p_1\}\\
X_6 &=\{L_3,p_1^2\}-\hbar^2\{\frac{7}{x},p_2\}+\hbar^2\{ \frac{-3 y}{x^2},p_1\}\\
X_7 &=p_1^3+\frac{\hbar^2}{2}\{ \frac{3}{x^2},p_1\}
\end{split}
\end{equation*}

 In this case we find nine
third order integrals, two of which are trivial ($H p_2$ and
$p_2^3$), and four are purely quantum integrals. In the classical
limit they correspond to integrals of the free motion. Only the
first three can be associated with the corresponding classical
integrals of $V=\frac{a}{x^2}$.

The most interesting potentials are obtained by setting all the
$A_{ijk}$ involved in \eqref{compatx} equal to $0$. The
expressions for $f_1$,$f_2$,$f_3$,$f_4$ greatly simplify and
equations \eqref{eqx1} to \eqref{eqx4} can be solved directly. The
nonlinear compatibility condition for these four equations reduces
to

\begin{equation}\label{elliptique}
\h ^2 V'(x)^2= 4 V(x)^3 +\a V(x)^2 +\b V(x) +\gamma,
\end{equation}

\noindent where the $\a$,$\b$,$\gamma$ are arbitrary real
integration constants. Equation \eqref{elliptique} is the
well-known equation for elliptic functions which can be rewritten
as

\begin{equation}\label{elliptique2}
\h ^2 V'(x)^2= 4 (V(x)-A_1)(V(x)-A_2)(V(x)-A_3).
\end{equation}

The constants $A_i$ are either all real, or one of them is real
and the other two are complex conjugated. If all three constants
are real, we obtain either finite or singular potentials of the
form

\begin{equation}
\begin{split}
V_1&=(\hbar \omega)^2 k^2 sn^2(\omega x,k),\\
V_2&=\frac{(\hbar\omega)^2}{ sn^2(\omega x,k)},
\end{split}
\end{equation}
respectively.

If we have e.g. $A_3=A_2^*$ and $Im A_2 \neq 0$, we obtain the
singular potential

\begin{equation*}
V_3=\frac{(\hbar\omega)^2}{2 (cn(\omega x,k)+1)}
\end{equation*}

\noindent (throughout we have $0\leq k \leq1$, $\w \in \bf{R}$).

The special cases with $k=0$ or $k=1$, which arise when two roots
coincide, can be expressed in terms of elementary functions. The
most interesting example is the "soliton" potential,

\begin{equation*}
V_{1a}=\frac{(\hbar\omega)^2}{\cosh^2(\omega x)},
\end{equation*}
 obtained
by setting $k=1$ in $V_1$. If we set $k=0$, or $k=1$ in $V_2$, we
get a singular periodic, or nonperiodic potential, respectively,
namely

\begin{equation*}
\begin{split}
V_{2a}=\frac{(\hbar\omega)^2}{\sin^2(\omega x)},\\
V_{2b}=\frac{(\hbar\omega)^2}{\sinh^2(\omega x)}.
\end{split}
\end{equation*}

For all these potentials $\w$ is an arbitrary constant, hence
there exist potentials of arbitrary amplitude for all nonzero
values of $\h$.

Finally, if all roots coincide, we reobtain the known
superintegrable potential

\begin{equation*}
V_{4}=\frac{\hbar^2}{x^2},
\end{equation*}

which explains the extra integrals found previously for that
potential.

The other potentials $V_1$,$V_2$,$V_3$ also satisfy

\begin{equation*}
\frac{\h^2}{4 } \frac{V_{xxx}}{V_x}-3V=\a,~~~~~\a=A_1+A_2+A_3,
\end{equation*}
\noindent a consequence of \eqref{elliptique}.

The two nontrivial integrals of motion for all these potentials
can be written as

\begin{equation}
\begin{split}
X_1&=\{L_3,p_1^2\}+\{(\alpha - 3
V(x))y,p_1\}+\{-\a x+2xV(x)+\int V(x)dx,p_2\}\\
X_2&=p_1^3+\frac{1}{2}\{3 V(x)-\a,p_1\}.
\end{split}
\end{equation}

 The second integral can be
trivially obtained by the commutation of the first one with $p_2$.

%\section{Higher order symmetries}
%[[ceci illustre la necessite de produire une definition
%appropriee]]
%This particular form of an integral allows us to add
%to the potential any function of $y$ only. This way we loose $P_2$
%as an integral, but it was shown in \cite{STW} that a system
%allowing separation of variables in cartesian coordinates (i.e.
%$V(x,y)=V_x(x)+V_y(y)$)possesses second order symmetries of the
%form

%\begin{equation}
%  X_1=p_1^2+V_x(x)
%  X_2=p_2^2+V_y(y)
%\end{equation}

%This means that the potentials proportional to $\h^2$ we found,
%possessing a third order symmetry depending only on $x$ and $p_1$,
%can be modified by any term depending on $y$ and still have a
%third order symmetry, in addition with the second order ones. These integrals are related though by
%\begin{equation}
%[[a confirmer]]
%\end{equation}
%  X_1=p_1^2+V_x(x)
\section{Conclusion}

We have found all potentials in two-dimensional Euclidian space
$E_2$ that allow one first- and at least one third-order integral
of motion. In the classical case the result provides no new
superintegrable potentials; all the potentials found allow second
order integrals and the third order integrals are consequences of
the second order ones. In the case of quantum mechanics the result
is quite different. Any potential satisfying the elliptic function
equation \eqref{elliptique} will be superintegrable in the above
sense, i.e. will allow the first-order integral $p_2$ and two
nontrivial third order integrals.  All those "behave well" in the
classical limit, that is they are proportional to $\h^2$ and
therefore their classical limit is the (superintegrable) free
motion.

No new superintegrable systems are found for rotationally
invariant potentials $V(r)$, neither in the classical, nor in the
quantum case. Thus all potentials found above are of the form
$V=V(x)$, i.e. are actually one-dimensional. The problem however
remains two-dimensional as the kinetic energy and the integrals of
motion also involve the $y$ direction.

There is also an interesting link with soliton theory (\cite{AC}).
All new superintegrable potentials obtained above are also
translationally invariant solutions of the Korteweg-de Vries
equation. The same potentials occur in the rational, trigonometric
and elliptic Calogero-Moser-Sutherland models (\cite{DV}).

The difference between classical and quantum integrable and
superintegrable systems with higher order symmetries makes the
systematic search for such systems very interesting. First of all,
Drach's study of classical integrable systems should be completed.
His systems are really complex ones and most of them do not exist
in real Euclidian space. Moreover it is not clear how complete his
list is. On the other hand, Ra\~nada (\cite{Ra}) has shown that 7
out of 10 Drach potentials are "reducible" in the sense that they
are superintegrable and allow two second-order integrals. The
third order integral found by Drach is the Poisson commutator of
the second-order ones.

The problem of classifying quantum systems with third-order
integrals remains open and the conditions of Section
\ref{condition} provide the means for finding all such systems.

Work is in progress on superintegrable systems in two-dimensional
Euclidian space with one second and one third-order invariant, as
well as with two third order ones.

\section{Acknowledgements}
The authors thank M.B.Sheftel and P. Tempesta who participated in
the early stages of this project for interesting discussions. The
final version of this article was written while both authors were
visiting the Departemento de Fisica Teorica II of the Universidad
Complutense. They thank the Departemento and specially M.A.
Rodriguez for hospitality and helpful discussions. S.G. benefited
from an NSERC student fellowship. The research of P.W. was partly
supported by research grants from NSERC of Canada, FCAR du
Qu\'ebec and NATO.

\end{document}